\documentclass[conference]{IEEEtran}
\IEEEoverridecommandlockouts
\usepackage{graphicx}
\usepackage{url}
\usepackage{color}
\usepackage{cite}
\usepackage[bookmarks=false]{hyperref}
\usepackage{amsmath}
\usepackage{tikz}
\usepackage{paralist}
\usetikzlibrary{decorations.pathreplacing}

\begin{document}

\title{\Huge (Short Paper) Towards More Reliable\\ Bitcoin
Timestamps\vspace{-0.65cm}}

\author{\IEEEauthorblockN{Pawel Szalachowski}
\IEEEauthorblockA{SUTD\\pawel@sutd.edu.sg}
\thanks{This work was supported in part by the National Research Foundation
(NRF), Prime Minister's Office, Singapore, under its National Cybersecurity R\&D
Programme (Award No. NRF2016NCR-NCR002-028) and administered by the National
Cybersecurity R\&D Directorate.}
}

\maketitle

\begin{abstract}
    Bitcoin provides freshness properties by forming a blockchain where each
    block is associated with its timestamp and the previous block.  Due to these
    properties, the Bitcoin protocol is being used as a decentralized, trusted,
    and secure timestamping service.  Although Bitcoin participants which
    create new blocks cannot modify their order, they can manipulate timestamps
    almost undetected.  This undermines the Bitcoin protocol as a reliable
    timestamping service.  In particular, a newcomer that synchronizes the
    entire blockchain has a little guarantee about timestamps of all blocks. 

    In this paper, we present a simple yet powerful mechanism that increases
    the reliability of Bitcoin timestamps.  Our protocol can provide evidence
    that a block was created within a certain time range.  The protocol is
    efficient, backward compatible, and surprisingly, currently deployed SSL/TLS
    servers can act as reference time sources.  The protocol has many
    applications and can be used for detecting various attacks against the
    Bitcoin protocol.
\end{abstract}

\section{Introduction}
\label{sec:intro}
Bitcoin~\cite{nakamoto2008bitcoin} is a cryptocurrency successful beyond all
expectations.  As a consequence of this success and properties of Bitcoin,
developers and researchers try to reuse the Bitcoin infrastructure to build new
or enhance existing systems.  One class of such systems is a decentralized
timestamping service.  For instance, the OpenTimestamps
project~\cite{ots} aims to standardize
blockchain time\-stamping, where a timestamp authority, known from the previous
proposals~\cite{rfc3161}, is replaced by a blockchain.  Other, more focused
applications that rely on the blockchain timestamps include trusted
record-keeping service~\cite{lemieux2016trusting,gao2017decentralized},
decentralized audit systems~\cite{liwill,spoke2015blockchain}, document signing
infrastructures~\cite{jamthagen2016blockchain}, timestamped
commitments~\cite{clark2012commitcoin}, or secure off-line payment
systems~\cite{dmitrienko2017secure}.  Reliable timestamps are also vital for
preventing various attacks against the Bitcoin protocol. For instance,
Heilman proposed a scheme~\cite{heilman2014one} which with unforgeable
timestamps can protect from the selfish mining
strategy~\cite{eyal2014majority}.

By design, the Bitcoin protocol preserves the order of events (i.e., \textit{weak
freshness}), however, accurate time of events (i.e., \textit{strong freshness})
is questionable, despite the fact that each block has a timestamp associated.
In practice, Bitcoin timestamps can differ in hours from the time maintained by
Bitcoin participants (\textit{nodes}), and in theory can differ radically from
the actual time (i.e., time outside the Bitcoin network).  Effectively, the
accurate time cannot be determined from the protocol, which limits capabilities
of the Bitcoin protocol as a timestamping service, and which impacts the
security of the protocol~\cite{gervais2015tampering}.

In this work, we propose a new mechanism for improving the security of Bitcoin
timestamps.  In our protocol, external timestamp authorities can be used to
assert a block creation time, instead of solely trusting timestamps put by block
creators.  Our scheme is efficient, simple, practical, does not require any
additional infrastructure nor any changes to the Bitcoin protocol, thus can be
deployed today.  Interestingly, currently existing SSL/TLS servers can
act as time authorities.

\section{Background and Preliminaries}
\label{sec:pre}
\subsection{Freshness in the Bitcoin Blockchain}
Bitcoin is an open cryptocurrency and transaction system.  Each transaction is
announced to the Bitcoin network, where nodes called \textit{miners} collect and
validate all non-included transactions and try to create (\textit{mine}) a
\textit{block} of transactions by solving a cryptographic proof-of-work puzzle,
whose difficulty is set such that a new block is mined about every 10 minutes.
Each block has a \textit{header} that contains the block's metadata.
Transactions are represented as leaves of a Merkle tree~\cite{Merkle1988} whose
root is included in the \textit{header};  hence, with the header, it is possible
to prove that a transaction is part of the given block.  Every block header
contains also a field with a hash of the previous header to link the blocks
together.  Due to this link, the blocks create an append-only
\textit{blockchain}.  Additionally, headers include Unix timestamps that
describe when the corresponding block was mined.  These timestamps are used as
an input for proof-of-work puzzles and are designed to impede an adversary from
manipulating the blockchain.

Freshness properties offered by the Bitcoin protocol are unclear.  Since the
blockchain is append-only, weak freshness is provided by design (i.e.,
blocks are ordered in the chronological order). Timestamps associated with
blocks are validated in a special way. Namely, a node considers a new block's
timestamp $T$ as valid if:
\begin{compactenum}
    \item $T > $ the median timestamp of previous eleven blocks, and
    \item $T - 2h <$ \textit{network time} (defined as the median of the
        timestamps returned by all nodes connected to the node).
\end{compactenum}

Each node maintains its local Bitcoin timer, which is defined as the node's
local system time plus the difference between this time and the network time.
However, the timer cannot be adjusted more than 70 minutes from the local system
time.

As it is not required that all nodes have accurate time, timestamps encoded in
headers may not be even in order, and their accuracy is estimated to hours.
Manipulation of the Bitcoin network time is possible and can result in severe
attacks~\cite{culubas}.
Furthermore, as Bitcoin timestamps depend only on time of nodes, timestamps can
differ radically from the actual time, outside the network.  Another issue is
that nodes synchronizing the entire blockchain have hardly any guarantees
about the previous blocks' creation times.  Given that, it is clear that the
Bitcoin protocol does not provide strong freshness,  what limits the Bitcoin
blockchain applicability for time-sensitive applications (like accurate
timestamping). 

\subsection{Timestamping Service}
\label{sec:pre:time_srv}
The time-stamp protocol (TSP)~\cite{rfc3161} is a standard timestamping protocol
built on top of the X.509 public key infrastructure (PKI).  In the protocol,
a client that wishes to timestamp data contacts a timestamp authority (TSA)
with the data's hash.  The TSA signs the hash along with the current timestamp
and returns the signed message to the client.  The message, with the TSA's
certificate and the data, allows everyone to verify that the data was
timestamped at the given time. 

For simple description, we present our protocol as compliant with TSP. However,
with minor or no changes, our scheme can be combined with other services, like
currently existing PKIs or secure time synchronization services (see
\autoref{sec:impl:time}).

\subsection{System Model}
Our protocol introduces the two following parties:
\begin{compactitem}
    \item\textbf{Timestamping authority (TSA)} runs a service that timestamps
        documents according to the TSP protocol presented in
        \autoref{sec:pre:time_srv}.
    \item\textbf{Verifier} is an entity that wants to verify when a new
        (upcoming) blockchain's block was mined. A verifier can interact with
        the Bitcoin network by reading blocks and sending transactions and can
        interact with a (chosen) trusted TSA.
\end{compactitem}
We assume that the used cryptographic primitives are secure.  We assume an
adversary able to mine Bitcoin blocks, and her goal is to introduce a new block
with an incorrect timestamp (i.e., deviating from the TSA's time) undetected. 

\subsection{Notation}
Throughout the paper we use the following notation:

\begin{tabular}{l p{6cm}}
$\{msg\}_A$ & denotes the message $msg$ signed by $A$, \\
$h(.)$ & is a cryptographic hash function, \\
$\|$ & is the string concatenation, \\
$r\xleftarrow{R}S$ & denotes that $r$ is an element randomly selected from the set $S$,\\
$\{0,1\}^n$ & is a set of all $n$-bit long strings,\\
$B_i$ & denotes the $i$th blockchain's block,\\
$H_i$ & denotes the $i$th block header,\\
$T_x$ & is a Unix timestamp expressed in seconds. \\
\end{tabular}

\section{Description of the Protocol}
\label{sec:details}
\subsection{High-Level Overview}
\begin{figure}[h!]
  \centering
  \includegraphics[width=0.97\linewidth]{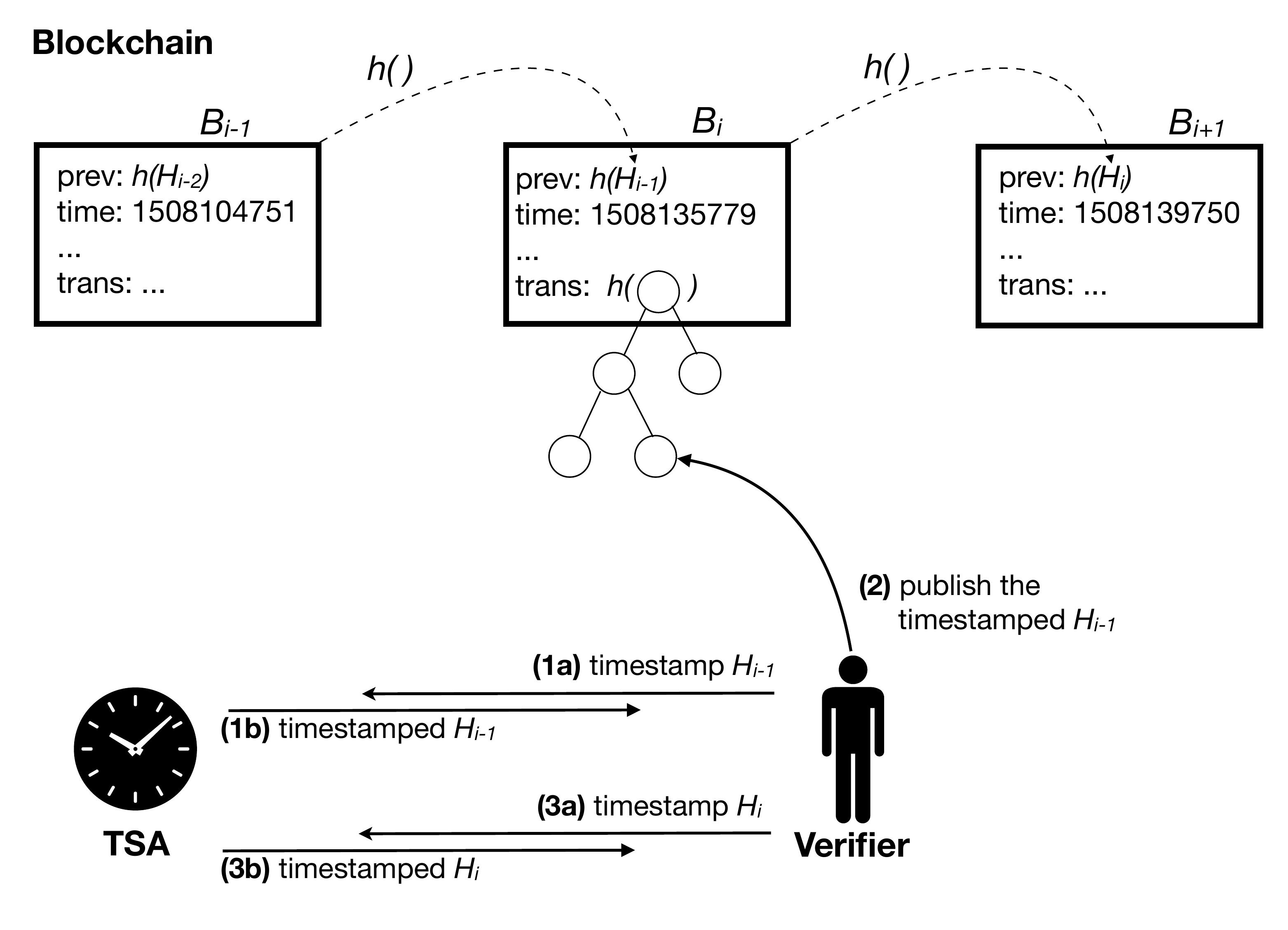}
  \caption{A high-level overview of the protocol.}
  \label{fig:overview}
\end{figure}

The main idea behind our scheme is to combine an external TSA with the
blockchain, such that a verifier can create a cryptographically-provable series
of events that asserts when a given block was mined (i.e., when all transactions
associated with the block were published).  A simplified description of our
protocol is presented in \autoref{fig:overview}.

The protocol starts, when a verifier sees a new block $B_{i-1}$.  The verifier
extracts the block header $H_{i-1}$ and contacts a TSA to timestamp $H_{i-1}$.
Then, the TSA returns a timestamped and signed $H_{i-1}$ to the verifier.  This
message states that the block $B_{i-1}$ is older than the message itself (i.e.,
than its timestamp).  Next, the verifier publishes the timestamped and signed
message in the blockchain.  The corresponding transaction is published in the
subsequent block $B_i$.  As the transaction is included in the block, it implies
that the block is newer than the transaction (i.e., the block is newer than the
timestamp associated with the transaction).  Finally, the verifier extracts the
header $H_i$ of this block and timestamps it with the TSA.  Now, the verifier
has evidence that the block $B_i$ was created between the timestamped
messages (i.e., between their timestamps).

\subsection{Details}
As presented above the verifier interacts with the TSA and the Bitcoin network.
Everyone can act as a verifier, and TSAs can be chosen arbitrarily by verifiers.
The protocol is initiated independently by a verifier by executing the
following:
\begin{compactenum}
    \item On receiving the $(i-1)$th block $B_{i-1}$ with the block header
        $H_{i-1}$, the verifier:
    \begin{compactenum}
        \item selects a random value $R_0 \xleftarrow{R} \{0,1\}^n$,
        \item prepares data $D_0 = h(R_0\|H_{i-1})$ to be timestamped.
    \end{compactenum}
    \item The verifier contacts a TSA to timestamp $D_0$.
    \item The TSA returns a timestamped and signed message $\{D_0,T_0\}_{TSA}$.
    \item On receiving this message the verifier:
		\begin{compactenum}
            \item computes $C = h(\{D_0, T_0\}_{TSA})$ as a commitment,
            \item encodes $C$ within a Bitcoin transaction, and
            \item propagates the transaction across the network, such that it
                is included in the subsequent block $B_i$.
		\end{compactenum}
    \item On receiving the $i$th block $B_i$, with the block header $H_{i}$, the
        verifier:
    \begin{compactenum}
        \item selects a random value $R_1 \xleftarrow{R} \{0,1\}^n$,
        \item prepares data $D_1 = h(R_1\|H_{i})$ to be timestamped,
        \item creates $P_C$ as a Merkle tree inclusion proof of the
            transaction containing $C$.
    \end{compactenum}
    \item The verifier contacts the TSA to timestamp $D_1$.
    \item The TSA returns a signed message $\{D_1, T_1\}_{TSA}$.
    \item Now, the verifier has the following information
        \begin{equation}
            \label{eq:proof}
            R_0, H_{i-1}, \{D_0,T_0\}_{TSA}, R_1, H_i, \{D_1, T_1\}_{TSA},P_C,
        \end{equation}
        which constitutes a \textit{proof} that the block $B_i$ was mined
        between $T_0$ and $T_1$. 
    \item To verify whether the block has a correct timestamp, the verifier
        checks if the following is satisfied:
        \begin{equation*}
            T_0 < H_i\text{'s timestamp} < T_1.
        \end{equation*}

\end{compactenum}
\medskip

The verifier can terminate the protocol at the step 9. However, to verify the
creation time of the subsequent block, he can compute a new commitment $C =
h(\{D_1, T_1\}_{TSA})$ and conduct the protocol from the step 4b onwards.

For the sake of a simple presentation, we include $D_0$ and $D_1$ in the proof
(see \autoref{eq:proof}), but they are redundant as can be computed from $R_0,
H_{i-1}$ and $R_1, H_i$, correspondingly.  We also describe the protocol with a
single TSA.  However, it is easy to extend the scheme to multiple TSAs. In such
a case, the verifier timestamps the $D_0$ and $D_1$ messages with multiple TSAs,
and computes the commitment $C$ as a hash over the TSAs messages corresponding
to $D_0$.

\section{Analysis}
\label{sec:analysis}
First, we claim that the verifier executing the protocol obtains a provable
series of events that given block was mined in a given time range.  Hence, an
adversary cannot introduce a block with an invalid timestamp undetected.
(Although we present our protocol in the adversarial setting, invalid timestamps
can be introduced by benign miners with desynchronized clocks.)

\begin{figure}[b!]
\centering
	\begin{tikzpicture}[scale=1]
	\node[align=center] at (7.5,-0.225) {\textit{time}};
	\node[align=center] at (1,1) {block $B_{i-1}$\\is created};
	\draw [thick,->] (1,0.55) -- (1,0.15);
	\node[align=center] at (3.25,0.6) {mining period};
	\node[align=center] at (5.25,1) {block $B_i$\\is created};
	\draw [thick,->] (5.25,0.55) -- (5.25,0.15);
	\draw [thick,decorate,decoration={brace,amplitude=6pt,raise=0pt}] (1.25,0.15) -- (5,0.15);
	\draw [thick,->] (0,0) -- (8,0);
	\draw [thick,->] (1.6,-0.55) -- (1.6,-0.15);
	\node[align=center] at (1.6,-1) {timestamp $D_0$\\and publish $C$};
	\draw [thick,->] (5.6,-0.55) -- (5.6,-0.15);
	\node[align=center] at (5.6,-1) {timestamp $D_1$};
	\end{tikzpicture}
    \caption{Timeline of the events in the protocol.}
    \label{fig:time}
\end{figure}
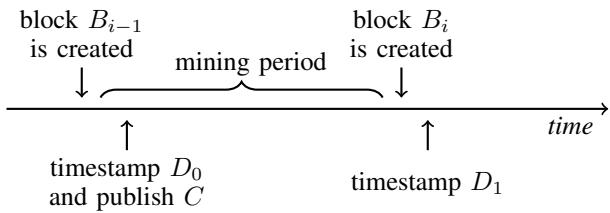

The timeline of the protocol events is presented in \autoref{fig:time}.  When
the verifier notices the block $B_{i-1}$ he creates $D_0$ by hashing a
random value $R_0$ and the block's header $H_{i-1}$. $D_0$ is timestamped by the
TSA, and the commitment $C$ is computed as a hash of this timestamped message.
With $C$ the verifier can check that it was indeed created after the $B_{i-1}$ as the
header of $B_{i-1}$ (i.e., $H_{i-1}$) was used to create it. Therefore, the
block $B_{i-1}$ is older than the timestamp $T_0$.  Then, the commitment is
propagated among the network and finally included in the newly created block
$B_i$.  The verifier, with the header $H_i$ of the new block can prove that
$C$ is part of this block (using the Merkle inclusion proof $P_C$), thus
it has to be older than the block.  Next, the verifier from a random value $R_1$
and the block's header $H_i$ creates $D_1$, which is timestamped by the TSA.
The message from the TSA ($\{D_1, T_1\}_{TSA}$) states that $D_1$ was
created before $T_1$, and because $D_1$ is derived from $H_i$, it implies that
the block $B_i$ was created before $T_1$.  Finally, the verifier equipped with
this information (see \autoref{eq:proof}) can check whether the block's
timestamp is correct (i.e., $\in[T_0,T_1]$).

Our protocol provides much better freshness properties than the Bitcoin protocol
alone.  As depicted in \autoref{fig:time}, if the verifier creates and publishes
$C$ immediately after the block $B_{i-1}$ is observed and timestamps $D_1$ after
the block $B_i$ is observed, then the accuracy of timestamping is approximately
equal the block creation time (currently, estimated as 10 minutes).  The
verifier can increase the accuracy by creating and publishing multiple
commitments in a sequence, such that the difference between timestamped $D_0$ and
$D_1$ decreases.

The protocol is described in the scenario where the commitment $C$ appears in
transactions corresponding to the block $B_{i}$.  Although the propagation in
the Bitcoin network is fast when compared to the average block creation
time~\cite{decker2013information}, it may happen that $C$ is included in a
later block. In such a case, our protocol still provides guarantees about the
blocks in between.  For example, if the commitment $C$ appears not in $B_i$
but in the block $B_{i+1}$, then the proof states that blocks $B_i$ and
$B_{i+1}$ were mined between $T_0$ and $T_1$.

The verifier generates a random value $R_0$ that together with $H_{i-1}$ is
timestamped by the TSA as $\{D_0, T_0\}_{TSA}$, which in turn is hashed into the
commitment $C$.  The commitment is published in the blockchain, however, $R_0$ is
not revealed.  This construction protects the protocol from censorship by an
adversary that wishes to manipulate the timestamp.  Without this random value,
the adversary could just keep timestamping hash of $H_{i-1}$ every second,
learn all possible commitments for the block header, and censor the verifier's
transaction.  With a large random value (e.g., chosen from $\{0,1\}^{128}$),
generating all possible commitments is infeasible, hence the adversary cannot
distinguish between a regular transaction and the verifier's transaction.

Although we do not consider malicious TSAs, the protocol provides means to keep
them accountable.  If the TSA returns the signed messages such that $T_0>T_1$,
then the verifier has an evidence that the TSA misbehaved.  More specifically,
the verifier can show that $D_0$ is older than $D_1$ (by showing that $D_1$ was
created using $H_i$, which contains $C$ created from $D_0$ which was timestamped
at $T_0$), which proves that the TSA contradicted itself.  Moreover, the TSA
does not know secret random values $R_0$, $R_1$, hence cannot learn what is
being timestamped. (However, colluding TSA and adversary could censor
commitments.)

\section{Practical Considerations}
\label{sec:impl}
\subsection{Commitments Encoding}
In our protocol, a verifier publishes commitments in the blockchain (see the step
4c of the protocol).  This message is computed as a hash thus is short and can
be encoded on the blockchain in many ways.  One way is to publish a transaction
with the commitment encoded within the 20-byte long \textit{receiver of
transaction} (\texttt{pay-to-pubkey-hash}) field.  An alternative could be to
encode messages into other fields or to use the \texttt{OP\_RETURN}
instruction~\cite{shirriff2014hidden}.  

Storing non-transaction data in the Bitcoin blockchain is regarded by many
members of the Bitcoin community as a spam or even a vandalism.  We agree that
using the Bitcoin blockchain as a highly distributed database negatively
influences its performance.  However, we believe that our protocol will be seen
as a positive contribution to the ecosystem, as firstly, it aims to improve
the security of the protocol, and secondly, the overhead introduced is marginal.
Moreover, this overhead can be minimized by publishing commitments through a
system like OpenTimestamp, which aggregates and publishes data in the blockchain
efficiently.

\subsection{Timestamping Service}
\label{sec:impl:time}
We describe our protocol to be compliant with the timestamping service as
defined in the RFC 3161~\cite{rfc3161} (see \autoref{sec:pre:time_srv}).  There
are many providers of this service, both commercial and free. However, our
protocol, with minimal or no changes, can be combined with other currently
existing infrastructures.

Surprisingly, today's SSL/TLS servers can act in our protocol as TSAs.  The
SSL/TLS protocol supports Diffie-Hellman (DH) as a key-exchange algorithm.  In
such a case, a server sends to a client the \texttt{ServerKeyExchange} message,
that among other parameters, signs the DH parameters, and client's and server's
random values.  These random values start with a timestamp field, hence it is
possible to timestamp a document by the server's key by setting the client's
random value to a document's hash~\cite{edstrom2012blog}.  As, SSL/TLS is
becoming ubiquitous and the DH exchange is widely
supported~\cite{adrian2015imperfect,szalachowski2018blockchain}, web servers of reputable organizations
(e.g., \texttt{mozilla.org}) or high-profile websites (like \texttt{google.com}
or \texttt{live.com}) can be used as TSAs.

Another infrastructure that with minimal changes can implement the TSA
functionality is secure time synchronization infrastructure. For instance,
Roughtime~\cite{roughtime}, a recent proposal by Google, provides signed
timestamps.   To prevent replay attacks, a client inputs its nonce which
together with a timestamp is signed by the server.  To implement the TSA
functionality, a client just inputs $D_{i-1}$ as a nonce, like in the protocol.
One small change is caused by the design of Roughtime where, for efficiency
reasons,  servers sign responses in batches.  Hence, values returned by servers
are encoded differently, however still are verifiable and can be used
analogically as the TSA's output from the protocol (see the steps 3 and 7).

\section{Conclusions}
\label{sec:conclusions}
In this paper, we presented a method of strengthening the reliability of Bitcoin
timestamps.  Our protocol is efficient, backward compatible, and can provide
much stronger freshness guarantees than the Bitcoin protocol alone.  Our method
can be combined with currently existing and widespread security infrastructures
like the SSL/TLS PKI.  Although we presented our scheme in the Bitcoin context,
it is also applicable to other blockchain-based platforms.

The protocol can be deployed in many applications.  Verifiers can run the
protocol to detect misbehaving nodes. The protocol can be part of a detection
system against time-related attacks or can be combined with a system like
OpenTimestamps to enhance it.  Proofs can be also publicly published, so nodes
that in the future download and validate the entire blockchain will have much
better assurance about the event timeline.

\bibliographystyle{abbrv}
\bibliography{ref,rfc}
\end{document}